\documentstyle[preprint,prl,aps]{revtex}
\newcommand{\be}{\begin{equation}}
\newcommand{\ee}{\end{equation}}
\newcommand{\ben}{\begin{eqnarray}}
\newcommand{\een}{\end{eqnarray}}

\newcommand{\til}{\tilde}

\newcommand{\ra}{ \rangle}
\newcommand{\la}{ \langle}
\newcommand{\h}{{\cal{H}}}

\newcommand{\kf}{|f \rangle}
\newcommand{\brf}{\langle f|}
\newcommand{\kfN}{|f_{N} \rangle}

\newcommand{\kfNj}{|f_{\Nj} \rangle}

\newcommand{\al}{\alpha }
\newcommand{\kaln}{|\alpha_n \ra}
\newcommand{\baln}{\la \alpha_n|}

\newcommand{\kalnt}{|\til{\alpha}_n \ra}
\newcommand{\balnt}{\la \til{\alpha}_n|}

\newcommand{\kalkkkts}{|\til{\alpha}_{k+1}^{k+1} \ra}

\newcommand{\kt}{|t \ra}
\newcommand{\bt}{\la t |}

\newcommand{\sumni}{\sum_{n=1}^\infty}
\newcommand{\sumn}{\sum_{n=1}^N}
\newcommand{\sumnj}{\sum_{n=1 \atop n \ne j}^N}
\newcommand{\sumnl}{\sum_{l=1 \atop l \ne j}^N}

\newcommand{\sep}{\;\; \;\;\;;\;\;\;\;\;}
\newcommand{\spa}{\, ; \,}

\newcommand{\sumkk}{\sum_{n=1}^{k+1}}
\newcommand{\kalnkts}{|\til{\alpha}_n^k \ra}
\newcommand{\balnkts}{\la \til{\alpha}_n^k|}
\newcommand{\kalnkkts}{|\til{\alpha}_n^{k+1} \ra}
\newcommand{\balnkkts}{\la \til{\alpha}_n^{k+1}|}

\newcommand{\OPKK}{\hat{P}_{V_{k+1}}}
\newcommand{\OPKKT}{{\OPKK}^\dagger}

\newcommand{\kpsik}{|{\psi}_{k+1} \ra}
\newcommand{\bpsik}{\la {\psi}_{k+1} |}
\newcommand{\kpsikp}{\frac{\kpsik}{||\kpsik||^2}}

\newcommand{\psit}{\til{\psi}}

\newcommand{\kpsitn}{|{\psit}_{n} \ra}
\newcommand{\bpsitn}{\la {\psit}_{n}|}
\newcommand{\kpsitjf}{|{\psit}_{j}^f \ra}
\newcommand{\bpsitjf}{\la {\psit}_{j}^f|}
\newcommand{\kpsijf}{|{\psi}_{j}^f \ra}
\newcommand{\bpsijf}{\la {\psi}_{j}^f|}
\newcommand{\spann}{{\mbox{span}}}

\newcommand{\Nj}{N/j}
\newcommand{\vn}{V_{N}}
\newcommand{\vnj}{V_{N/\alpha_j}}
\newcommand{\vnjc}{V_{N/\alpha_j}^\bot}
\newcommand{\pvn}{\hat{P}_{V_N}}
\newcommand{\pvnj}{\hat{P}_{\vnj}}
\newcommand{\pvnjc}{\hat{P}_{\vnjc}}
\newcommand{\kg}{|g\ra}
 
\newcommand{\kgc}{|g^\bot\ra}

\newcommand{\altt}{\til{\alpha}}

\newcommand{\kalj}{{|\alpha}_j\ra}
\newcommand{\balj}{{\la\alpha}_j|}
\newcommand{\kaltjN}{|\altt_j^N \ra}
\newcommand{\baltjN}{\la{\altt}_j^N |}

\newcommand{\kaltnN}{|\til{\alpha}_n^N \ra}
\newcommand{\baltnN}{\la \til{\alpha}_n^N |}
\newcommand{\kaltnNj}{|\til{\alpha}_n^{\Nj} \ra}
\newcommand{\baltnNj}{\la \til{\alpha}_n^{\Nj}|}

\newcommand{\subnj}{ \;;\;n=1,\ldots, j-1, j+1, \ldots,N}

\newcommand{\ii}{\hat{I}}
\newcommand{\intt}{\lim_{T \to \infty} \int_{-T}^{T}}

\begin{document}


\title{Backward Adaptive Biorthogonalization }

\author{L. Rebollo-Neira} \address{
NCRG, Aston University\\ 
http://www.ncrg.aston.ac.uk\\ 
Birmingham B4 7ET, United Kingdom}
\draft

\maketitle

\begin{abstract}
A backward biorthogonalization approach is proposed,  which
modifies biorthogonal functions so as to generate orthogonal
projections onto a reduced subspace. The technique is relevant
to problems amenable to be represented by a general linear model.
In particular, problems of data compression, noise reduction and sparse
representations may  be tackled  by the proposed approach.
\end{abstract}
\section{Introduction}
We introduce a backward biorthogonalization technique relevant to 
the general linear model. Any  data which 
may be described in terms of a linear combination of waveforms
satisfies this model.
For instance, the response of a physical system to a particular
interaction varying as a function of a parameter, say $t$,
is often represented by a measurable quantity $f(t)$, which 
is amenable
to be expressed in the fashion:
\be 
f(t)= \sum_{n}^N c_n^N \alpha_n(t), 
\label{mo}
\ee 
where the model waveforms $\alpha_n \spa n=1,\ldots,N$ are derivable by 
recourse to physical considerations. The determination 
of the coefficients $c_n^N \spa n=1,\ldots,N$ 
entails to solve the inverse problem when the 
function $f(t)$ is measured. 
The superscript $N$ indicates that, 
unless the waveforms $\alpha_n$ are orthogonal, the  
appropriate coefficients  $c_n^N $ depend on the order of the model, i.e., 
the  number $N$ of waveforms 
$\alpha_n$ being considered in (\ref{mo}). 
If such waveforms are linearly independent, then, there exits a set 
of reciprocal functions  $\altt_n^N \spa n=1,\ldots ,N$
which is biorthogonal to the former, i.e.
$\la \altt_n^N | \alpha_m \ra = \delta_{n,m}$ \cite{yo,rs}. 
Here the superscript $N$ 
indicates that the biorthogonal  functions $\altt_n^N$ allow for 
constructing orthogonal projections onto the subspace $V_N$ 
spanned by $N$ waveforms $\alpha_n$.  
Hence, the coefficients $c_n^N$ of the linear expansion (\ref{mo}) 
approximating a function $f$ at best (in a minimum distance sense)
can be obtained by computing the inner products 
$c_n^N = \la \altt_n^N | f \ra$ \cite{rs,relo}.\\
Since the reciprocal set $\altt_n^N \spa n=1,\ldots,N$  
(and therefore the coefficients $c_n^N$) depend on 
the number $N$ of total waveforms, they should be 
recalculated when this number is enlarged or reduced. 
This feature of non-orthogonal expansions is 
discussed in \cite{rs,relo,iee}, where a 
recursive methodology is introduced for transforming the 
reciprocal set
$\altt_n^N \spa n=1,\ldots,N$ into a new one 
 $\altt_n^{N+1} \spa n=1,\ldots,N+1$. The latter is guaranteed to yield 
orthogonal projections onto the subspace $V_{N+1}$ 
 arising  by the inclusion of 
a waveform $\alpha_{N+1}$ in $V_{N}$, i.e., 
$V_{N+1}= V_N  + \alpha_{N+1}$.\\ 
Here we wish to consider the converse situation: Let us 
suppose that the reciprocal waveforms $\altt_n^N \spa n=1,\ldots,N$ 
are known  and we want to modify them so as to obtain orthogonal  
projections onto a subspace $\vnj$ arising by eliminating an 
element, say the $j$-th one, from $V_N$. Then,  
$\vnj=\spann\{\al_n \spa n=1,\ldots, j-1, j+1,\ldots,N\}$.  
Our aim is to construct the corresponding reciprocal functions 
$\altt_n^{\Nj} \spa n=1,\ldots, j-1, j+1,\ldots,N$ by 
modifying the previous $\altt_n^N \spa n=1,\ldots,N$.\\
Let us suppose that in the summation of (\ref{mo}) we want to retain 
only some terms and approximate $f(t)$ by a linear combination of those elements. 
Thus, to obtain the best approximation of $f(t)$  by a 
linear combination of such a nature, we need to recalculate the coefficients 
corresponding to the waveforms we wish to retain. If we simply disregarded the
coefficients of the unwanted terms, but did not recalculate the 
remaining ones, 
the approximation would not be optimal in a minimum distance sense.
The approach we propose in this Communication  allows 
for the necessary modifications of coefficients so as to achieve 
the optimal approximation. The method is based on an iterative technique 
capable of adapting biorthogonal functions in order to generate 
orthogonal projections onto a reduced subspace.\\
The paper is organised as follow: Sections II introduces the 
notation, discusses the motivation to the proposed approach and
summarises a previously introduced forward biorthogonalization 
method \cite{rs,relo}. 
Section III discusses the proposed biorthoganization technique 
to transform biorthogonal functions in order to build orthogonal 
projections onto a reduced subspace. The conclusions are drawn in Section IV.
\section{Notation, Background and Motivation to the approach}
Adopting Dirac's vector notation \cite{di} we represent an
element $f$ of a Hilbert space $\h$ as a vector $\kf$ and
its dual as $\brf$. Given
a set of $\delta$-normalized continuous orthogonal vectors
$\{ \kt \,;\, -\infty < t < \infty \,;\, \bt t' \ra= \delta(t-t')\}$,
the unity operator in $\h$ is expressed
\be
{\ii}_{\h}= \intt \kt  \bt \; dt.
\label{i1}
\ee
Thus, for all $\kf$ and $|g \ra \in \h$, by inserting ${\ii}_{\h}$ in
$\brf  g \ra$, i.e,
\be
\brf  {\ii}_{\h} |g \ra= \intt \brf t \ra \bt g \ra\; dt
\ee
one is led to a representation of $\h$ in terms of
the space of square integrable functions,
with $\bt g \ra= g(t)$ and $\la g \kt= { \bt g \ra}^\ast= {g^\ast(t)}$,
where ${g^\ast(t)}$ indicates the complex conjugate of $g(t)$.\\
Let vectors $\kaln \in \h \; ;\;  n=1,\ldots, \infty$ be a Riesz basis
for $\h$.  Hence, all $\kf \in \h$  
can be expressed as the linear span
\be
\kf = \sumni c_n \kaln 
\label{moca}
\ee
and there exists a reciprocal basis
$\kalnt \; ;\;  n=1,\ldots, \infty$ for $\h$ to which the former basis is
biorthogonal i.e., $\balnt \alpha_m \ra= \delta_{n,m}$ \cite{yo}. 
The reciprocal basis allows to
compute the coefficients $c_n$ in (\ref{moca}) as the
inner products               
\be                          
c_n =  \balnt  f \ra= \intt {\altt_n^\ast(t)} f(t) \,dt.
\label{cn}                   
\ee
Thus, 
\be
\kf = \sumni  \kaln \balnt f \ra 
\ee  
so that, by denoting 
\be
\hat{I}= \sumni \kaln \balnt,
\label{io}
\ee
(\ref{moca})  can be recast as $f = \hat I f$, which implies
that $\hat I$ is a representation of the identity operator in $\h$
and we have the following generalization of the Plancherel-Parseval 
identity
\be
||\kf||^2 = \la f |\hat{I}| f \ra= \sumni \til{c}_n^\ast c_n 
\label{pa}
\ee
with  ${c}_n$ as in (\ref{cn}) and  $\til{c}_n^\ast = \la f \kaln$. \\ 
If the basis 
$\kaln \spa n=1,\ldots ,\infty$ is orthogonalized  and we denote
by  $ \kpsitn \spa n=1,\ldots ,\infty$ the corresponding
orthogonal  vectors after normalization to unity, then the new basis
is self-reciprocal, i.e.,  it  satisfies the orthonormality
condition $\la \psit_m| \psit_n \ra = \delta_{m,n}$ and  provides a
representation for the identity operator as given by
\be
\hat I = \sumni \kpsitn  \bpsitn.
\label{ino}
\ee
This representation of the identity operator
can be seen as a particular case
of (\ref{io}),  by considering the basis and its reciprocal
identical to  $\kpsitn \spa n=1,\ldots ,\infty$. 
The equivalence between
(\ref{io}) and (\ref{ino}) holds only  when  both 
sums run to infinity.
Because, on the one hand if the sum in (\ref{ino}) is truncated up to
$N$ terms we obtain an operator, $\hat P$, given by
\be   
\hat P =  \sumn  \kpsitn  \bpsitn,
\ee  
which is the orthogonal projector onto the subspace $V_N$ 
spanned by  $N$  vectors $\kaln \spa n=1,\ldots ,N$. On the other hand,
by truncating (\ref{io}) up to $N$ terms one obtains an
operator
\be   
\hat Q= \sumn  \kaln \balnt,
\ee   
which is idempotent, and hence a projector, but as it
fails to be self-adjoint   
it is not an orthogonal projector operator.
As a consequence, the approximation of $\kf$ that
we obtain by truncating the expansion (\ref{mo}) up to $N$
terms is {\underline{not}} the best approximation of $\kf$ that can be
obtained by a linear superposition of $N$  vectors  $\kaln$.
If one wishes for orthogonal
projections by means of biorthogonal families, then
biorthogonal vectors $|\altt_n^N  \ra \spa n=1,\ldots,N$
specially devised for such a purpose must be constructed.
The superscript $N$ indicates that if the subspace $V_N$ is
enlarged (or reduced) each function should be recalculated. 
\subsection{Forward Adaptive Biorthogonalization}  
Let $\kaln$ be a set of linearly independent vectors  
and let vectors $|\psi_n \ra$ be obtained by orthogonalizing 
the formers in such a way that $|\psi_n\ra = \kaln - \hat{P}_{V_{n-1}}\kaln$, 
where $\hat{P}_{V_{n-1}}$ is the 
orthogonal projector operator onto the subspace $V_{n-1}$ spanned  
by $ |\al_l \ra \spa l=1,\ldots, n-1$. 
Then, it is proved in \cite{rs,relo} that 
vectors $\kalnkkts$ arising from $|\psi_1 \ra = |\alpha_1 \ra $ through the
recursive equation:
\ben
\kalnkkts &=&\kalnkts -
{\kpsikp}\la \alpha_{k+1} \kalnkts
\sep n=1,\ldots,k\nonumber\\
\kalkkkts &=& \kpsikp
\label{reca}
\een
are biorthogonal to vectors $\kaln \;;\; n=1,\ldots,k+1$
and provide a representation of the orthogonal projection operator
onto $V_{k+1}$ i.e., 
\be
\OPKK = \sumkk \kaln \balnkkts= \OPKKT =\sumkk \kalnkkts \baln.
\label{teu}
\ee 
As discussed in \cite{iee}, in order to reduce 
numerical errors the vectors $|\psi_k\ra $ are conveniently 
computed by Modified Gram Schmidt procedure  or 
Modified Gram Schmidt with pivoting \cite{iee,rise}.\\  
Since the unique vector in $V_{k+1}$ minimizing 
the distance to an arbitrary vector $ \kf \in \h$ is obtained by the
operation $\OPKK f$ \cite{rs,relo}, it follows from (\ref{teu}) that the 
coefficients $c_n^{k+1}$ of the linear expansion
\be
\sum_{n=1}^{k+1} c_n^{k+1} \kaln
\label{le}
\ee
which approximates an arbitrary  $\kf \in \h$ at best
in a minimum distance sense, can be recursively obtained
as:
\ben
c_n^{k+1}&=& c_n^{k} - \balnkts \alpha_{k+1}\ra
\frac{\bpsik f \ra}{||  \kpsik ||^2} \sep n=1,\ldots,k \nonumber\\
c_{k+1}^{k+1}&=&\frac{\bpsik f\ra}{||\kpsik||^2}, \label{c2}
\een
with
$c_1^{1}= \frac{\la \alpha_{l_1}| f \ra}{|||\alpha_{l_1}\ra||^2}$.\\
This technique, yielding forward approximations, has been shown to be 
of assistance in sparse signal representation by waveforms selection 
\cite{relo} as well as data set selection \cite{pre}. Nevertheless, 
in those and other application areas, it is clear the 
need for a technique yielding approximations in the 
opposite direction.  Hence the motivation to the approach of 
the next section.
\section{Backward Adaptive Biorthogonalization}
Let $\vnj$ denote the subspace which is left by removing
the vector $\kalj$ from $\vn$, i.e,
\be
\vnj= \spann \{|\alpha_1 \ra,  \ldots, |\alpha_{j-1} \ra, |\alpha_{j+1} \ra,
\ldots, |\alpha_{N} \ra\}
\ee
and let $|\altt_n^{\Nj}\ra \spa n=1,\ldots, j-1, j+1, \ldots,N$ be the
corresponding reciprocal family which allows to express
the orthogonal projector operator onto $\vnj$ as
\be
\pvnj =\sumnj \kaln \baltnNj = \sumnj \kaltnNj \baln.
\label{pvnj}
\ee
Assuming that the biorthogonal vectors $\kaltnN \spa n=1,\ldots, N$
yielding a representation of $\pvn$ as given by
\be
\pvn =\sumn \kaln \baltnN = \sumn \kaltnN \baln 
\label{pvn}
\ee
are known, 
our goal is to modify such vectors so as to
obtain  the corresponding set
$|\altt_n^{Nj}\ra \spa n=1,\ldots, j-1, j+1, \ldots,N$ yielding $\pvnj$
as in (\ref{pvnj}).\\
We start by writing
\be
\pvn = \pvnj + \pvnjc,
\label{pde}
\ee
where $\pvnjc$ is the orthogonal projector onto $\vnjc$,
the orthogonal complement of $\vnj$ in $\vn$. Thus,
$\vnjc$ contains only one linear independent vector,
arising by subtracting from $\kalj$ its component in $\vnjc$,
i.e.,
\be
\pvnjc= \kpsitjf \bpsitjf
\label{psif}
\ee
where
\be
\kpsijf= \kalj - \pvnj \kalj
\ee
and $\kpsitjf= \frac{\kpsijf}{||\kpsijf||}$.\\
(Note: we use the notation $\kpsijf$ to differentiate this
fresh vector from the previous $|\psi_j \ra$ introduced in Section II.A)\\
Using now (\ref{pvnj}), (\ref{pvn}) and  (\ref{psif}) we
express (\ref{pde}) as
\be
\sum_{n=1}^N \kaln \baltnN = \sumnj \kaln \baltnNj + \kpsitjf \bpsitjf.
\label{pdee}
\ee
Taking the inner product of both sides of (\ref{pdee}) with
$\bpsitjf$, and using the fact that $\bpsitjf \alpha_n \ra= 0 $ for
$n\neq j$, we obtain:
\be
\bpsitjf \alpha_j\ra \baltjN= \bpsitjf.
\label{fre}
\ee
Moreover, since 
 $\bpsitjf \alpha_j\ra= \la \alpha_j  | \alpha_j  \ra
- \la \alpha_j | \pvnj | \alpha_j \ra = \bpsijf  \psi_j^f \ra$,  
it follows from (\ref{fre}) that 
$||\bpsijf||= ||{\baltjN}||^{-1}$. Hence,  
vector $\bpsitjf$ turns out to be
\be
\bpsitjf= \frac{\baltjN}{||\baltjN||}.
\label{psitf}
\ee
Taking now the inner product of both sides of (\ref{pdee}) with
every $\baltnN \spa n=1,\ldots,j-1, j+1,\ldots,N$ we
obtain the equation we wanted to find:
\be
\baltnNj = \baltnN - \la \altt_n^N \kpsitjf \bpsitjf  
\sep n=1,\ldots, j-1, j+1, \ldots,N.
\label{rec}
\ee
The following theorem demonstrates that the modification of
vectors $\kaltnN$ as prescribed in (\ref{rec}) provides us with
biorthogonal vectors $\kaltnNj \spa n=1,\ldots, j-1, j+1,\ldots,N$
 rendering orthogonal
projections.\\
{\bf{Theorem 1:}} Given a set of vectors $\kaltnN \spa n=1,\ldots, N$
biorthogonal to vectors $\kaln \spa n=1,\ldots, N$ and
yielding a representation of $\pvn$ as given in (\ref{pvn}),
a new set of biorthogonal vectors
$\kaltnN \spa n=1,\ldots, j-1, j+1, \ldots,N$
yielding a representation for $\pvnj$, as given in (\ref{pvnj}), 
can be obtained from the following equations
\be
\kaltnNj = \kaltnN - \frac{\kaltjN \baltjN \altt_n^N \ra}{||\kaltjN||^2} 
\sep  n=1,\ldots, j-1, j+1, \ldots,N.
\label{rec2}
\ee
{\it{Proof:}} Let us first use (\ref{rec2}) to write
\be
\pvnj= \sumnj \kaln \baltnNj= \sumnj \kaln \baltnN  -
 \sumnj  \kaln  \frac{\la \altt_n^N \kaltjN \baltjN}{||\kaltjN||^2}.
\label{pvnje}
\ee
To prove that (\ref{pvnje}) is the orthogonal projector onto
$\vnj$ we show that a) $\pvnj \kg = \kg$ for all $\kg \in \vnj$
and b) $\pvnj \kgc = 0$ for all  $\kgc$ in the orthogonal complement of
$\vnj$ in $\h$. \\
Indeed, every  $\kg \in \vnj$ can be expressed as a
linear combination $\kg = \sumnj a_n \kaln$, for some coefficients $a_n
\spa n=1,\ldots, j-1, j+1, \ldots,N$ and since, by hypothesis,
$\baltnN \alpha_l \ra  = \delta_{n,l}$ from (\ref{pvnje})
we have:
\be
\pvnj \kg = \sumnj  \sumnl \kaln a_l \baltnN  \alpha_l \ra =
\sumnj a_n  \kaln = \kg,
\ee
which proves a).\\
To prove b) we write $\sumnj  \kaln \baltnN = \sumn \kaln \baltnN -  \kalj \baltjN=
 \pvn - \kalj \baltjN$
and recast (\ref{pvnje}) as
\ben
\pvnj &=& \pvn - \kalj \baltjN -
 \frac{ \pvn  \kaltjN \baltjN}{||\kaltjN||^2}
+\frac{ \kalj \la \altt_j^N \kaltjN \baltjN}{||\kaltjN||^2} \nonumber\\
&=& \pvn - \frac{ \pvn  \kaltjN \baltjN}{||\kaltjN||^2}.
\label{sim}
\een
Now, since $\la \alpha_n  \kgc= \delta_{n,j}
\la \alpha_j \kgc \spa n= 1,\ldots,N$,  it
follows that  $\pvn \kgc =  \kaltjN \balj  g^\bot \ra $ and,
since $\kaltnN \in \vn$,  it follows that  $\baltjN g^\bot \ra = \baltjN  \pvn \kgc =
\la \altt_j^N \kaltjN  \balj  g^\bot \ra.$ Hence,
\ben
\pvnj \kgc &=& \pvn  \kgc -  \frac{\kaltjN \baltjN \pvn \kgc}{||\kaltjN||^2}
\nonumber \\
 &=& \kaltjN \la \alpha_j \kgc - \kaltjN \la \alpha_j \kgc= 0. 
\een
The biorthogonality property of vectors $\kaltnNj \subnj$ is an immediate 
consequence of the  biorthogonality property of vectors $\kaltnN$, 
as readily follows by taking the inner product of 
both sides of (\ref{rec2}) with each vector $\baln \subnj$.\\ 
Since $\pvnj= \sumnj  \kaln \baltnN$ has been proved to be a projector, 
it is self-adjoint. Hence (\ref{pvnj}) holds $\;\Box$\\
{\bf{Corollary 1:}} Let $\kfN$ be the orthogonal
projection of and arbitrary  $\kf \in \h$ onto $\vn$, 
i.e
\be
\kfN = \pvn \kf =\sum_{n=1}^{N} c_n^{N} \kaln
\ee
with $c_n^N= \baltnN f \ra \spa n=1,\ldots,N$ assumed to be known.
Hence, the coefficients $c_n^{\Nj}$  of the
orthogonal projection of $\kf$ onto $\vnj$
are obtained from the known coefficients $c_n^N$ as follows:
\be
c_n^{\Nj}=c_n^{N} - \frac{\la \altt_n^N \kaltjN c_j^N}{||\kaltjN||^2}.
\label{core}
\ee
The proof trivially stems from (\ref{pvnje}), since  
$\pvnj \kf  =\sumnj c_n^{\Nj} \kaln$ implies 
$c_n^{\Nj}= \baltnNj f\ra  \;\;\Box$\\
{\bf{Corollary 2:}} For $\kf \in \h$,  let 
$\kfN$ be as above and $\kfNj=\pvnj \kf$. Then, the following 
relation between $||\kfN||$ and $||\kfNj ||$ holds: 
\be 
||\kfNj||^2 = ||\kfN ||^2 - \frac{|c_j^N|^2}{||\kaltjN||^2}. 
\label{col}
\ee
{\it{Proof:}} Using (\ref{sim}) and the fact that projectors are 
 self-adjoint and idempotent, it follows that
\ben 
||\kfNj||^2 & = &  \brf \pvnj \kf = \brf \pvn \kf - 
       \frac{\la f \kaltjN \baltjN f \ra}{||\kaltjN||^2} \nonumber \\
       & = & ||\kfN ||^2 - \frac{|c_j^N|^2}{||\kaltjN||^2}\;\;\;\;\Box
\een
So far we have discussed how to modify the coefficients of a 
linear expansion when one of its  components is  removed. 
Nevertheless, we have given no specification on how to choose such an element. 
We are  now in a position to address this point,  since    
the last Corollary suggests how the selection  could be 
made optimal.  The following proposition is in order.\\
{\bf{Proposition 1:}} Let 
\be
\kfN= \pvn \kf  = \sumn c_n^N \kaln  
\label{ap1}
\ee
be given by the coefficients $c_n^N \spa n=1\ldots,N$, and 
let 
\be
\kfNj= \pvnj \kf  = \sumnj c_n^{\Nj} \kaln  
\ee 
be obtained by eliminating the  coefficient $c_j^N$  from (\ref{ap1}) 
and modifying  the remaining coefficients 
as prescribed in (\ref{core}).
The coefficient $c_j^N$ to be  removed as  
minimizing the norm of the 
residual error $|\Delta \ra =  \kfN -  \kfNj$ is the one 
yielding a minimum value of
\be 
\frac{|c_j^N|^2}{||\kaltjN||^2}. 
\label{crit}
\ee
{\it{Proof}}: Since $\pvn\pvnj=\pvnj \pvn=\pvnj$  we have: 
\be
||\kfN -  \kfNj||^2 = \brf \pvn \kf - \brf \pvnj \kf = ||\kfN ||^2 - || \kfNj||^2.
\ee
Hence, making use of (\ref{col}), we further have
\be
||\kfN -  \kfNj||^2 = \frac{|c_j^N|^2}{||\kaltjN||^2},
\ee
from which we gather that  $||\kfN -  \kfNj||^2$ is 
minimum if $\frac{|c_j^N|^2}{||\kaltjN||^2}$ is minimum $\Box$\\
Proposition 1 is relevant to backward approximation of a signal,
a common procedure in compression and noise reduction techniques.
The goal being to  
shrink coefficients so as to have a more economical representation
and/or reduce spurious information (noise).
Successive applications of criterion (\ref{crit}) leads
to an algorithm for recursive coarser approximations.
Indeed, let us assume that at the first iteration we eliminate
the $j$th-term yielding a minimum of (\ref{crit}).
We then construct the new reciprocal vectors as prescribed in
(\ref{rec2}) and the corresponding new coefficients as
prescribed in (\ref{core}). We are thus in a position to
repeat the  process and obtain a coarser approximation
of the previous one. If we denote by $|f^{(k)}\ra$ the
approximation arising at the $k$-step, a common stopping
criterion would recommend to cease the iteration process when
the following situation is reached:
\be
||| \kf \ra -   |f^{(k)} \ra||^2 > \delta, 
\label{crit2}
\ee
where $\delta$ is estimated according to the
desired  precision.
If the aim is to denoise a signal the value of $\delta$ may be set as the
variance of the noise, when available.
It is appropriate to remark, however, that in the context of some 
applications the selection criterion (\ref{crit}) may not be 
the adequate one.
Instead, other criteria based of statistical properties
may be required \cite{deno1,deno2,deno3}. In any case, regardless of the
criterion for selecting the coefficient $c_j^N$ to be overlooked,
if one wishes the
remaining ones to yield the optimal approximation in a 
minimum distance sense,
such coefficients should be modified as indicated
in (\ref{core}).
We illustrate next, by a simple example, the gain that results  
in following this prescription.\\
Let us consider $N=13$
elements $|\alpha_n\ra \spa n=1,\ldots,13$
whose functional representation are given by the following
shifted Mexican hat wavelet 
\ben
\alpha_{2n}(t)&=&  \la t |\alpha_{2n}\ra =
 A_{2n} e^{-(t+2n)^2} (1-(t+2n)^2) \nonumber\\
\alpha_{2n +1}(t)&= &\la t |\alpha_{2n+1}\ra = 
 A_{2n+1} e^{-(t- 2n - 1)^2}
(1-(t-2n-1)^2) \sep n=0,\ldots,6
\label{mexset}
\een 
where each $A_n$ is a constant which normalizes the corresponding 
function $|\alpha_{n}\ra$ to unity in the interval $[-4,4]$.  
We construct the biorthogonal functions 
 $\altt_n^{13}(t)$ (assumed to be known) by applying the forward biorthogonalization technique of 
Section II.A, but they could be constructed by any other available method.\\  
The signal $f(t)$ is considered to be the  pick plotted by the 
continuous line of Figure 1. Such signal is also expressible by    
a linear combination of the waveforms 
given in (\ref{mexset}). A high quality fitting results 
by using the corresponding 13 coefficients $c_n^{13}$, each of which is 
calculated as $c_n^{13} =\la \altt_n^{13} |f \ra$.  The 
actual numbers being the following: \\
\vspace*{0.05cm}\\
$2.8273 \;\; 2.4954 \;\; 2.4954 \;\; 1.9988 \;\; 1.9988 \;\;1.4989 \;\;1.4989$
\vspace*{0.05cm}\\
$0.8630 \;\;  0.8630\;\; 0.2957 \;\; 0.2957 \;\; 0.0648 \;\; 0.0648$\\
\vspace*{0.05cm}\\
We now disregard the last two coefficients, the ones of smallest 
magnitude, and use the remaining ones without modification. 
Although the neglected coefficients are quite small in comparison with 
some of the others, the approximation that results, 
represented by the  dotted line in Figure 1, does not fit correctly
the distribution tails. 
Nevertheless, if we disregard 
the same coefficients but modify the others by applying 
(\ref{core}) twice, the resulting approximation happens to coincide 
with the continuous line of Figure 1. To magnify the effect we 
wish to show, let us now disregard two more coefficients, those 
of value $0.2957$. The approximation that results from a simple 
truncation is shown by the darker dotted line of Figure 2. The slender 
dotted line plots our approximation. This very simple 
example clearly illustrates the  significance of the proposed  
modification of coefficients.
\section{Conclusions}
A recursive approach for adapting biorthogonal functions so as to 
obtain orthogonal projections onto a reduced subspace has been proposed. 
The required modifications are simple and easy to implement. 
The modified functions are used to adapt coefficients of a 
lower order linear model, in order to obtain an optimal approximation 
in a minimum distance sense. \\
A criterion for disregarding coefficients has being discussed. Such 
criterion leads to an iterative procedure for successive 
backward approximations which yields, at each iteration, minimal residual norm.
It should be stressed that, regardless of the criterion used for 
neglecting coefficients, the proposed 
approach may be applied to guarantee  optimality (in a minimum distance sense)
of the remaining approximation. We believe, thereby,  that this 
technique is potentially applicable to a broad range of problems including 
data compression, noise reduction and sparse representation. 
\section*{Acknowledgements}
Support from EPSRC (GR$/$R86355$/$01) is acknowledged.
I wish to thank  Dr S. Jain for corrections to the manuscript.

\newpage

\begin{figure}[h]
\begin{center}
\input{fittex.tex}
\vspace{4cm}\\
Figure 1: The continuous line represents a signal $f(t)$, which is
also expressible as a linear combination of the 13 waveforms given in
(\ref{mexset}). The dotted line is the approximation arising by
disregarding tow coefficients in such linear expansion. Our approach
coincides with the continuous line.
\end{center}
\end{figure}

\newpage

\begin{figure}[h]
\begin{center}
\input{fit2tex.tex}
\vspace{4cm}\\
Figure 2: Here the darker dotted line is obtained by disregarding four
coefficients. Our approach is represented by the slender dotted line.
\end{center}
\end{figure}

\newpage

{\bf Figure Captions}\\

{\bf Figure 1:} The continuous line represents a signal 
$f(t)$, which is
also expressible as a linear combination of the 13 waveforms given in
(\ref{mexset}). The  dotted line is the approximation arising by
disregarding tow coefficients in such linear expansion. Our approach
coincides with the continuous line.
            
{\bf Figure 2:} Here the darker dotted line is obtained by 
disregarding four
coefficients. Our approach is represented by the 
slander dotted line.
            
\end{document}